# Band-Gap Control *via* Structural and Chemical Tuning of Transition Metal Perovskite Chalcogenides

*Shanyuan Niu, Huaixun Huyan, Yang Liu, Matthew Yeung, Kevin Ye, Louis Blankemeier, Thomas Orvis, Debarghya Sarkar, David J. Singh, Rehan Kapadia, Jayakanth Ravichandran\**

Shanyuan Niu, Huaixun Huyan, Yang Liu, Kevin Ye, Thomas Orvis, Dr. Jayakanth Ravichandran
Department of Materials Science, Mork Family, University of Southern California
925 Bloom Walk, HED 216, Los Angeles, CA 90089, USA

Matthew Yeung, Louis Blankemeier, Debarghya Sarkar, Dr. Rehan Kapadia
Ming Hsieh Department of Electrical Engineering, University of Southern California
3740 McClintock Avenue, EEB 102, Los Angeles, CA 90089, USA

Prof. David J. Singh
Department of Physics and Astronomy, University of Missouri
Columbia MO 65211-7010, USA

Rational design or discovery of new materials, especially semiconductors, has been a key contributor to several electronic, photonic, and energy technologies. Among current semiconductors, the dominant materials such as silicon, III-V, II-VI are typically constructed by four-fold coordinated tetrahedral network of covalent bonds. Perovskites, a class of materials with highly symmetric close packed structure, have been researched heavily for their versatility in chemistry and physical properties. The organic-inorganic lead halide perovskites, which are shown to be efficient photovoltaic materials with laboratory scale power conversion efficiency of 17.9%,[1-7] provide an exciting example. However, its toxicity and instability issues pose serious challenges for practical applications. On the other hand, the well known perovskite oxides with chemical formula $ABO_3$, where A is an alkali, alkaline or rare earth metal and B is a transition metal, form the basis for our understanding of electronic properties of inorganic perovskites. Large energy differences between transition metal

*d*-orbital conduction band and O 2*p* orbital valence band in these materials typically lead to very large energy gaps (> 3 eV) rendering them unsuitable for solar energy conversion or visible optoelectronic applications. A new class of semiconductors, transition metal perovskite chalcogenides (TMPCs), addresses this deficiency. By replacing O with S or Se, we move the valence band composed of mainly chalcogen (S, Se) 3*p* or 4*p* orbitals higher and decrease the band gaps to Visible – IR range.

The maximum achievable power conversion efficiency of a single junction solar cell is sensitively related to the semiconducting material's band gap. Further, tunable band gap in a class of materials with similar structure and chemistry can also be of immense value for tandem solar cell architectures. TMPCs with a general formula, $ABX_3$, where A is Ba, Sr or Ca, B is Ti, Zr or Hf, and X is S or Se, are predicted to provide a platform for band-gap engineering in the visible – infrared region.[8] The ternary structure of perovskites opens up a larger design space compared to currently dominant single element (*e.g.* Silicon) and binary compound (*e.g.* GaAs, InP) systems. Theoretical calculations have shown that band gaps of individual TMPCs span from the Far IR to Visible spectrum with more than 10 materials falling in between 1.0 eV to 2.5 eV.[8] Although alloying has been successfully employed to tune band gaps for other semiconductor materials, substitutional defects can lead to disorder and inferior physical properties. Hence, the chemical and structural flexibility of the ternary structure in TMPCs can lead to wide range of heterostructures with good interfacial quality, which is especially attractive for various optoelectronic applications. Additionally, all the elements involved in TMPCs are benign and earth abundant, which allows TMPCs to address the sustainability issues clouding various advanced materials. In the case of $BaZrS_3$ (BZS) and $SrZrS_3$ (SZS), Ba, Sr, Zr and S has an earth abundance of 425, 370, 165 and 350 ppm, while Cd, Te, Sn and In has an abundance of 0.15, 0.001, 2.3 and 0.25 ppm respectively.[9] As inorganic compounds with close packed structure, these materials are also generally stable in ambient and elevated temperatures (up to ~400°C). We monitored the stability of these materials over the course of a year *via* samples stored in a normal desiccator and/or in air, and we measured no discernable degradation in structural and chemical characterizations, and physical properties. Another key advantage of TMPCs over conventional semiconductors is the large density of states of the conduction band. Large density of states leads to large absorption coefficients near the band edge (~ $10^5$ cm$^{-1}$),[10-12] and small absorption length (~ 100 nm absorbs >95% of incident solar photons above the band gap) in photovoltaic devices. Thin absorption layers increase the efficiency of collecting the excited carriers at the electrodes, and also lead to shorter synthesis times and lower materials costs.

Recently several theoretical calculations were reported on the electronic structure and optical properties of TMPCs. The electronic structure of BZS has been theoretically calculated and direct band gap of 1.7eV,[8,10,13] 1.72eV,[14] and 1.82eV[15] were reported. SZS is expected to show a band gap of 1.2 eV for needle-like phase (α-SZS) and 2 eV for distorted perovskite phase (β –SZS).[8] On the other hand, several TMPCs have been synthesized in ceramic form over the last half century, by either heating binary sulfide mixture for several weeks,[11,16-21] or sulfur substitution of corresponding oxides with $CS_2$[22,23] or $H_2S$.[17] Despite these numerous attempts, there is very little experimental data on their optical properties, as most reports focused on structural characterization. Recently, two groups reported optical properties of BZS. Meng et al.[14] reported a band gap of 1.85 eV for BZS synthesized by conventional solid-state reaction of binary mixtures (BaS and $ZrS_2$) with repeated annealing. Perera et al.[13] reported a band gap value of 1.7 eV with photoluminescence (PL) and diffuse reflectance measurements on BZS synthesized by high temperature sulfurization of oxides with $CS_2$. With these preliminary reports in mind, it is imperative to study the physical properties, especially optical properties, of TMPCs beyond BZS in depth to evaluate their potential as a new class of versatile semiconductors for optoelectronic applications.

In this manuscript, we will present a comprehensive study of three TMPCs (BZS, α-SZS, and β–SZS) synthesized by a novel, catalytic synthesis procedure, with in depth structural and chemical characterization. The vibrational and optical properties of the materials were studied by room temperature Raman spectroscopy, PL spectroscopy, and diffuse reflectance and transmittance measurements. We also evaluate the potential of TMPCs as a photovoltaic material through quantitative photoluminescence measurements, allowing extraction of the quasi Fermi level splitting with supporting first principles studies of their optical and electronic properties.

The synthesis method is based on conventional solid-state reaction of alkaline earth metal sulfides (BaS and SrS), and elemental sources (Zr and S) *with iodine as the catalyst*. Iodine has been used as a transporting agent in chemical vapor transport growth of single crystals. It greatly enhances the reactivity of transition metals by creating volatile compounds of the cationic species, such as low melting point iodides. Thus, it enables *one shot synthesis* of TMPCs within a few hours to days compared to other methods, which last for several weeks with repeated grinding and annealing for homogenization. Synthesized materials appeared in three different colors, as can be seen in the optical images in **Figure 1**, indicating distinct optical properties of each phase in the visible spectrum. Despite short reaction times, our characterizations showed very high structural and chemical quality of the final products. Powder XRD patterns showed single-phase samples for each material. Rietveld analysis showed lattice constants of a = 8.504 Å, b = 3.820 Å, and c = 13.917 Å for *α*-SZS; a = 7.103 Å, b = 9.758 Å and c = 6.731 Å for *β*-SZS; and a = 7.061 Å, b = 9.977 Å, and c = 7.014 Å for BZS, with a space group of *Pnma* (62) for all three materials. Further details on the refined structural parameters are included in the supporting information. Raman spectroscopy was performed at room temperature to gain further insight into the vibrational modes of the materials.

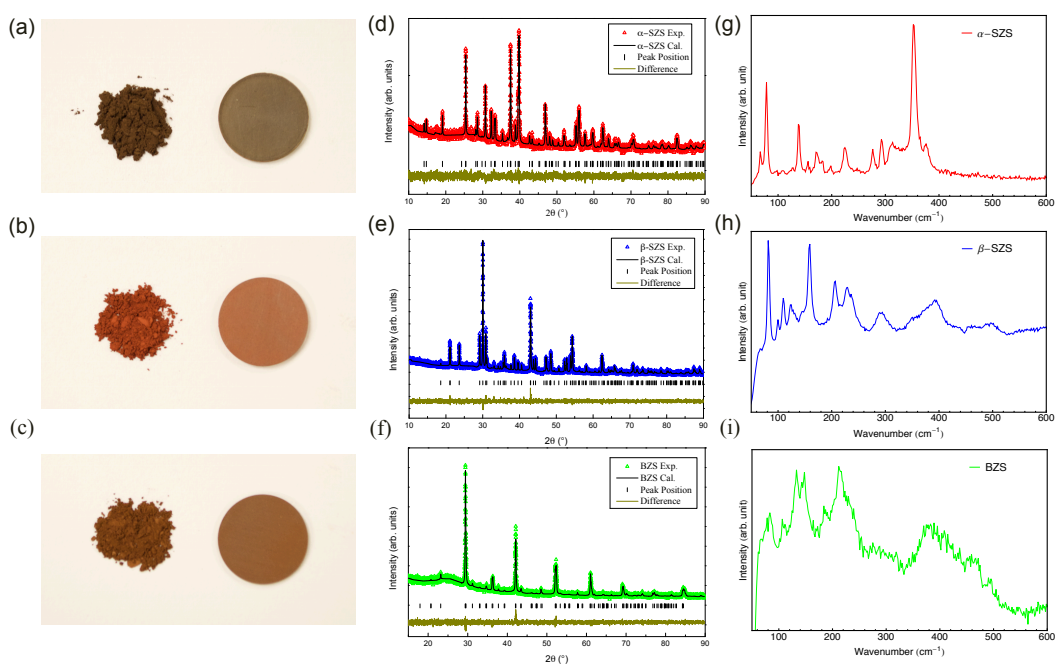

**Figure 1.** Optical images of synthesized powders and pelletized samples for (a) $\alpha$-SrZrS$_3$, (b) $\beta$-SrZrS$_3$ and (c) BaZrS$_3$. Plots of Powder XRD patterns with Rietveld analysis for (d) $\alpha$-SrZrS$_3$ (red), (e) $\beta$-SrZrS$_3$ (blue) and (f) BaZrS$_3$ (green). The black lines are simulated intensity profiles for three materials. Raman spectra for (g) $\alpha$-SrZrS$_3$, (h) $\beta$-SrZrS$_3$, and (i) BaZrS$_3$ at room temperature.

Chemical composition analysis of these polycrystalline materials was carried out using energy dispersive analytical X-ray spectroscopy (EDAX) and wavelength dispersive X-ray fluorescence spectrometry (WDXRF). The obtained chemical compositions in micron sized areas using EDAX were deduced as Ba: Zr: S= 1: 0.93: 3.06 for BZS, Sr: Zr: S= 1: 0.94: 2.90 for $\alpha$-SZS and Sr: Zr: S= 1: 0.92: 2.99 for $\beta$-SZS (EDAX spectra shown in **Figure 2**). The multi-spot analysis was performed using bulk sensitive WDXRF and showed consistent stoichiometry through 10 different points on the pellet as shown in Figure 2. The atomic percent ratio variation for Zr: Sr and S: Sr were within 1% and 4% respectively for $\alpha$-SZS and $\beta$-SZS; The Zr: Ba and S: Ba variation are within 1% and 3% respectively for BZS. These studies establish that the synthesized materials are crystalline and homogenous both at the microscopic and macroscopic scales.

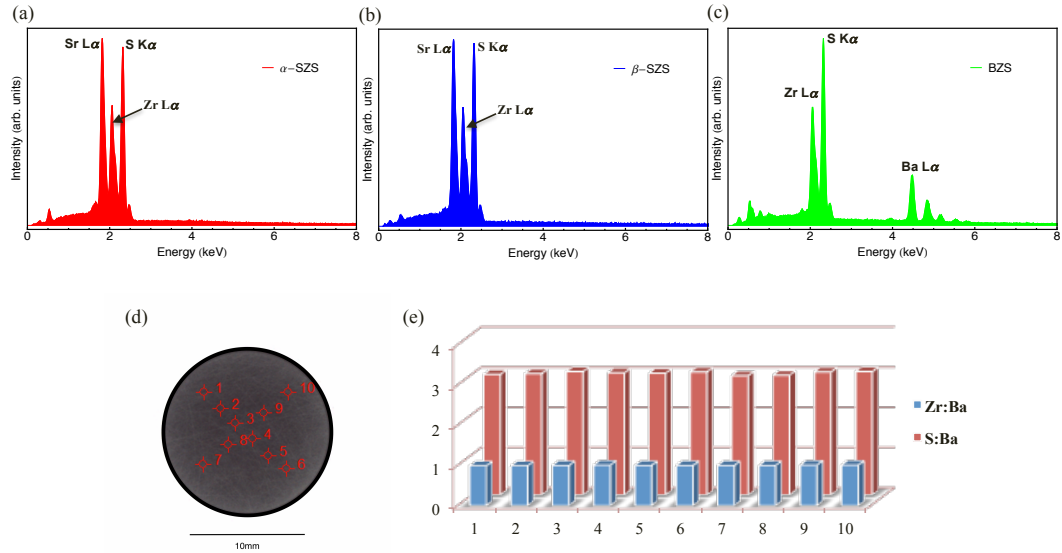

**Figure 2.** EDAX spectra for (a) α-SrZrS$_3$, (b) β-SrZrS$_3$ and (c) BaZrS$_3$. (d) Optical images of WDXRF mapping locations on BaZrS$_3$ pellet and (e) the corresponding stoichiometry ratios between Zr; Ba (blue) and S: Ba (red).

Optical properties of these materials were determined by room temperature PL. The PL peaks indicate band gap values of *α*-SZS as 1.53 eV, *β*-SZS as 2.13 eV and BZS as 1.81 eV, as shown in **Figure 3**. The theoretical values from calculations with the modified Becke Johnson (mBJ) potential are 1.12 eV, 1.73 eV and 1.55 eV for α-SZS, β-SZS and BZS, respectively. These values show the same trend as the experimental values, but are smaller by ~0.3 eV - 0.4 eV. All the compounds have direct gaps. The plots of band structure for all three compounds are shown in Figure S4 in supporting information. The shapes and positions of the valence and conduction band carrier pockets are illustrated in Figure 3(e). The intense PL peak of *β*-SZS was symmetric with FWHM of ~122 meV.

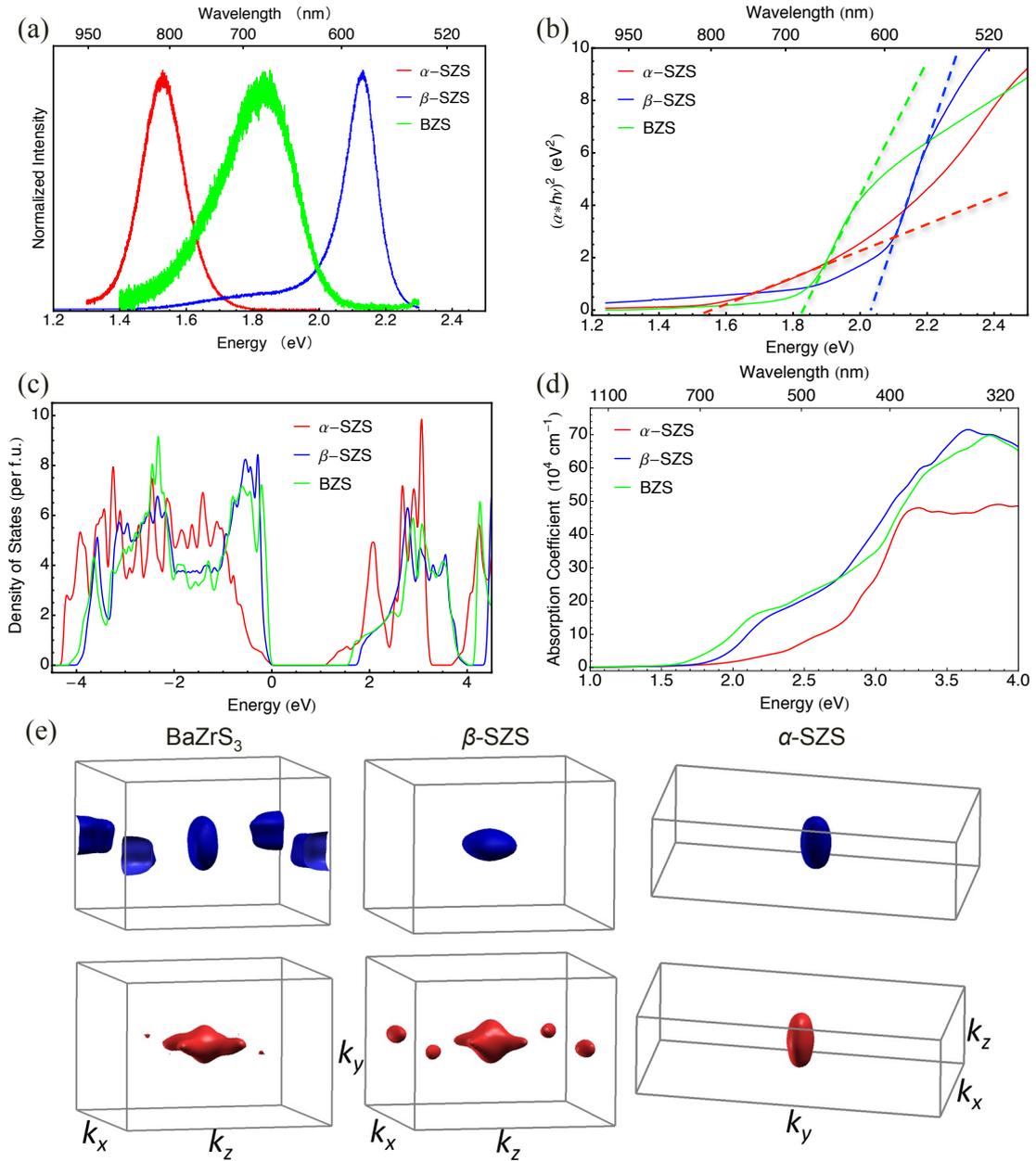

**Figure 3.** (a) PL spectra for *α*-SrZrS$_3$ (red), *β*-SrZrS$_3$ (blue), and BaZrS$_3$ (green) show band gap values of 1.53 eV, 2.13 eV and 1.81 eV respectively. (b) Band gap determination with absorbance value obtained from diffuse reflectance and transmittance measurements on translucent powder layer using Kubelka-Munk theory. The deduced band gap values are 1.52 eV for *α*-SrZrS$_3$ (red), 2.05 eV for *β*-SrZrS$_3$ (blue), and 1.83 eV for BaZrS$_3$ (green). (c) Calculated densities of states and (d) absorption spectra (direction averaged), and (e) isosurfaces of the calculated band structure 0.05 eV below the valence band maximum (VBM, blue) and 0.05 eV above the conduction band minimum (CBM, red) for three materials.

To quantify the intensity, the PL spectrum of *β*-SZS under same conditions was compared to single crystals of CdSe and InP, as shown in **Figure 4(a)**. *α*-SZS showed a symmetric but weaker peak with FWHM of 155meV, likely due to the lack of high symmetry corner sharing octahedral network. PL of BZS is not very symmetric or as intense as *β*-SZS, with a typical FWHM around 264 meV. The comparison of PL intensity for the three materials are shown in Figure S3 in supporting information. We are still investigating the trends in the luminescence yields for different TMPCs and how intrinsic and extrinsic factors affect them. The calculated optical absorption spectra of structurally similar *β*-SZS and BZS both show a particularly strong onset of absorption at the band edges, reaching values above $2\times10^5$ cm$^{-1}$ within a few tenths of eV above the gap. This is favorable for use as a solar absorber. It arises from the sharp onsets of the density of states (as shown in Figure 3(d)) at both the conduction and valence band edges, corresponding to weakly dispersive bands. Needle-like phase, *α*-SZS, shows a very different shape of the density of states, with broader bands at both the conduction and valence band edges and weaker onset in the absorption spectrum. Weak above gap absorption, despite a direct gap material, is a consequence of matrix elements and band structure effects. It is interesting to note that *α*-SZS and *β*-SZS with same chemical composition show very different band gap values and absorption spectra, indicating *structural control of optical properties*. The band gaps were also estimated by the diffuse reflectance and transmittance measurements on powder samples using a spectrometer with an integrating sphere to account for both absorption and scattering. The relationship between the measured diffusive reflectance and transmittance values to the absorption coefficient $k$, and a scattering coefficient $s$ is given by the Kulbeka-Munk theory.[24] The absorbance value ($\alpha = kd$) was used in $(\alpha \cdot h\nu)^2$ versus $h\nu$ plot to find the band gaps of the materials. Obtained values were 1.52 eV, 2.05 eV and 1.83 eV for *α*-SZS, *β*-SZS and BZS respectively, as shown in Figure 3(b). The diffuse reflectance measurements on infinitely thick powder layers were also done to extrapolate the band gaps, the values obtained match well with the translucent powder layers [see supporting information for more details].

Furthermore, it is worth noting that as seen in the iso-surface plots of Figure 3(e), the conduction bands of BSZ and *β*-SZS are highly corrugated, showing a four armed star shape with arms along the [101] directions. From the point of view of transport, when doped this type of shape combines parts of the Fermi surface that have effectively high velocities favorable for conduction, with other parts (along the arms) that are heavy and lead to high density of states. In the context of solar applications, this type of band structure favors the otherwise rare combination of high absorption (due to high density of states) and efficient electron collection. It also favors the combination of high thermopower and high conductivity in the context of thermoelectric materials.[25,26,27]

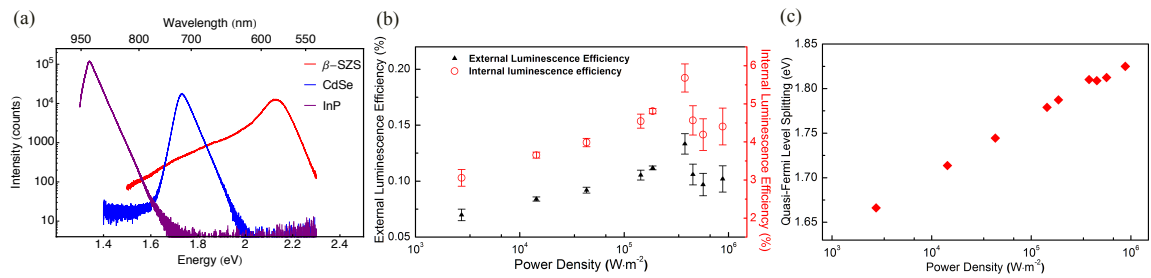

**Figure 4**. (a) PL intensity comparison of *β*-SrZrS$_3$ sample with InP and CdSe single crystal substrates. (b) External (black) and internal (red) luminescence efficiencies, and (c) Quasi-Fermi Level Splitting of *β*-SrZrS$_3$ as a function of incident power density.

In order to directly measure the potential of these materials for optoelectronic devices such as solar cells, we performed external luminescence efficiency ($\eta_{ext}$) measurements, and utilized results to extract the internal luminescence efficiency and quasi-Fermi level splitting ($\Delta E_F$) under illumination. These measurements were carried out by illuminating the sample with a known photon flux using a 532 nm laser, and then measuring the output photon flux utilizing a calibrated photoluminescence measurement system. Importantly, $\Delta E_F$ of a material defines the upper limits on the open circuit voltage from a solar cell,[28,29] and therefore the $\eta_{ext}$ measurement is a powerful, non-destructive measurement that allows determination of the quality of the measured material. Here, we utilize this technique to measure the quality of the grown *β*-SZS. Figure 5a shows a comparison of the photoluminescence curves of a reference single crystalline InP wafer, single crystalline CdSe wafer, and the *β*-SZS under the same illumination and measurement conditions. Importantly, we see that the photoluminescence intensity of the *β*-SZS is similar to the CdSe, and within an order of magnitude to the InP, which is promising, given that the grown material is polycrystalline while the references are single crystalline. Figure 5b shows both the measured $\eta_{ext}$ (black curve), and the internal luminescence efficiency (red curve), as a function of incident photon flux density. We see that the $\eta_{PL}$ varies from ~0.06% to 0.15% as the power density is increased to $2\times10^5$ W m$^{-2}$ and then drops down upon further increases in photon flux. This corresponds to an internal luminescence efficiency between 3%-6%, which is excellent for a polycrystalline material without any materials quality optimization. As a comparison, the $\eta_{ext}$ for world record polycrystalline CIGS and CdTe cells are between 0.0001-0.19%,[29] after decades of research and development. The overall shape of the $\eta_{PL}$ vs power density curve can be understood by considering the relative rates of radiative recombination, and non-radiative[30] Shockley-Read-Hall (SRH)[31] and Auger recombination.[31,32] Due to the carrier dependence of SRH, radiative, and Auger, which are proportional to the carrier density, the square, and the cube, respectively, it is expected that the measured efficiency will increase with incident optical density until it is limited by the non-radiative Auger recombination process.[32] This behavior is observed here, with the Auger recombination appearing to come into play around $10^6$ W m$^{-2}$, or 1000 suns. Figure 5c shows the extracted $\Delta E_F$ vs incident power density. The $V_{OC}$ corresponding to these would simply be $\Delta E_F/q$. While the bandgap for this specific TMPC is not optimal for a solar cell, the good optoelectronic quality measured, despite the polycrystalline nature and the lack of surface passivation suggests that epitaxial films of these materials may be excellent candidates for next-generation optoelectronic and energy devices.

In summary, TMPCs are proposed as a new class of semiconductors with great tunability and superior optoelectronic properties. Three representative examples, BZS and SZS in two room temperature stabilized phases, have been synthesized with an *iodine* catalyzed solid-state reaction process in sealed ampoules. Structural and chemical characterizations including XRD, Raman spectroscopy, EDAX and WDXRF establish high quality of the bulk ceramic samples, and the optimized synthetic procedure should be readily transferrable to other TMPCs. Intense room temperature PL signals indicate band gaps of 1.53 eV for α-SZS, 1.81 eV for BZS and 2.13 eV for β-SZS, and match well with the diffuse reflectance and transmittance measurements, and the theoretical calculations. Two structurally different SZS phases show distinct optical properties indicating structural control of optical properties. *α*-SZS shows weaker absorption with the smallest band gap, as compared to the sharp absorption onset and ultra-high absorption coefficients for the two higher symmetry phases, BZS and *β*-SZS, with absorption coefficients approaching $2\times10^5$ cm$^{-1}$ within only a few tenths of eV above the gap in the calculated absorption spectra. Additionally, the potential of these materials for solar energy harvest was evaluated by measurements of PL quantum efficiency and estimate of quasi Fermi Level splitting and compare well with existing high efficiency photovoltaic materials. These results indicate the excellent optical properties and broad chemical and structural tunability of TMPCs as promising candidates for photovoltaics, thermoelectrics and optoelectronic applications in general, and call for further in-depth experimental studies on TMPCs.

## Experimental Section

*Synthesis:* The starting materials, Barium Sulfide powder (Alfa Aesar 99.7%), Strontium Sulfide powder (Alfa Aesar, 99.9%), Zirconium powder (STREM, 99.5%), Sulfur pieces (Alfa Aesar 99.999%) and Iodine pieces (Alfa Aesar 99.99%) were stored and handled in an Argon-filled glove box. Stoichiometric quantities of precursor powders with a total weight of ~0.5 g was ground and loaded into a quartz tube with a diameter of 3/4" along with around 0.5 mg cm$^{-3}$ Iodine inside the glove box. The tube was then sealed using a blowtorch with oxygen and natural gas as the combustion mixture without exposing the contents of the ampoule to air. BZS, *α*-SZS and *β*-SZS samples were held at 600°C, 850°C, and 1100°C for 60 hours. All the samples were quenched to room temperature after the dwell time using a sliding furnace setup at approximately 100 K min$^{-1}$. Then the obtained samples were ground and pressed into 13 mm diameter pellets under uniaxial stress of around 600 MPa using a hydraulic cold press. The obtained pellets were mounted on acrylic epoxy base and polished with silicon carbide sand papers up to 12000 grit to yield a clean and reflective surface.

*Structural characterization:* The powder XRD characterization was carried out using a Bruker D8 Advance X-ray diffractometer with Co K$_\alpha$ radiation in Bragg-Brentano symmetric geometry with power setting of 35 kV and 40 mA. The sample stage was rotated at 15 rpm. Rietveld analysis were done using Fullprof Suite software on scans from 10 to 90 degrees with a 0.02 degree step and 5 seconds integration time for each data point. The details of the refining procedure and crystallite sizes estimate are discussed in detail in the supporting information.

*Analytical characterization:* The EDAX spectra were obtained in a JEOL 7001F analytical field emission scanning electron microscope equipped with energy dispersive X-ray spectrometer. All EDAX spectra were acquired with settings of 15 kV accelerating voltage, around 67 µA emission current and a working distance of 15 mm. Different magnifications were used from 100x to 5000x, the obtained chemical composition ratios were consistent. The spectra reported in this manuscript were recorded at 400x. The wavelength dispersive X-ray fluorescence spectrometry (WDXRF) multi-spot stoichiometry check was performed at 10 different spots on the pellets' surfaces with Rigaku's ZSX Primus IV.

*Optical and vibrational spectroscopy:* The room temperature PL measurements and Raman spectroscopy measurements were performed in a Renishaw inVia confocal Raman Microscope with a 532nm laser and a 100X objective lens. To eliminate the possible signals from surface contaminations, measurements were done right before and after sanding of the sample surface with sand paper. The results are consistent in both cases. The diffuse reflectance and transmission measurements were carried out in a Lambda 950 UV-VIS-NIR spectrometer with an InGaAs integrating sphere detector accessory. The scans were from performed for a wavelength range of 400 nm to 1000 nm, with 1 nm step size and repetition of 1 cycle. The detector switched from PMT tube to InGaAs at 900 nm, which gave rise to experimental artifacts in the form of jitters. The sample was a thin layer of powders, sandwiched in between two cover glass pieces. The diffuse transmittance $T$ was measured with samples clipped at the front entry port of the integrating sphere and the diffuse reflectance $R_0$ was measured at the exiting reflectance port with a black background. The transmittance and diffuse reflectance of two cover glass pieces were separately measured and accounted for in the final determination of the optical properties of the powder samples.

*Theoretical calculations:* We performed first principles calculations using the general potential linearized augmented planewave (LAPW) method[33] as implemented in the WIEN2K code. We used highly converged basis sets corresponding to a cut-off of $R_{min}k_{max}$=9.0, where $R_{min}$ is the minimum LAPW sphere radius and $k_{max}$ is the planewave sector cutoff. The LAPW sphere radii were 2.3 bohr for Zr and S, and 2.5 bohr for Ba and Sr. We did calculations with the modified Becke-Johnson potential (mBJ),[34] with experimental lattice parameters from literature and atomic positions determined by energy minimization with the PBE-GGA functional (see supplementary information). [35,36] We also performed calculations with the PBE+U method, applying U to the Zr d states as a parameter to adjust the gaps. We find that it is possible to obtain the experimental gaps in this way, although it requires the use of different values of the parameter U for each of the three compounds. Results shown are from the mBJ functional, without band gap adjustment, unless noted otherwise.

*Quantitative PL:* The quantitative photoluminescence measurements were performed in the same Renishaw inVia Microscope as the PL measurements. This tool was calibrated using a transfer standard InP wafer. This was carried out by first measuring the system instrument response function (IRF) with a broadband calibrated reference light source. Next, a reference InP single crystal wafer with a known quantum efficiency from an independently calibrated photoluminescence measurement system was then utilized to calibrate the absolute counts to photons conversion. The calibration was then cross checked across both tools.

**Supporting Information**

Supporting Information is available from the Wiley Online Library or from the author.

## Acknowledgements

This work was supported by a startup grant from USC Viterbi School of Engineering. The authors gratefully acknowledge the use of facilities at Dr. Brent Melot's Lab, Dr. Stephen Cronin's Lab and Center for Electron Microscopy and Microanalysis at University of Southern California for the results reported in this manuscript. The authors acknowledge the Rigaku Corporation for the WDXRF measurements. Work at the University of Missouri was supported by the Department of Energy, Basic Energy Sciences through the MAGICS center, award DE-SC0014607.